\documentclass[acmsmall,screen]{acmart}

\setcopyright{none}
\renewcommand\footnotetextcopyrightpermission[1]{}
\usepackage{minted}
\usepackage{wrapfig}
\usepackage{amsmath}
\usepackage{algorithmic}
\usepackage{graphicx}
\usepackage{textcomp}
\usepackage{xcolor}
\usepackage{booktabs}
\usepackage{multirow}
\usepackage{tcolorbox}
\usepackage{colortbl}
\usepackage{xspace}
\usepackage{listings}
\usepackage{adjustbox}
\usepackage{enumerate}
\usepackage{makecell}
\usepackage{pifont}
\usepackage[utf8]{inputenc}
\usepackage[T1]{fontenc}
\usepackage{microtype}
\usepackage{amsthm}
\usepackage{wasysym}
\usepackage{soul}
\usepackage{enumitem}
\usepackage{array}

\newcommand{\etal}{\hbox{\emph{et al.}}\xspace}
\newcommand{\eg}{\hbox{\emph{e.g.}}\xspace}
\newcommand{\ie}{\hbox{\emph{i.e.}}\xspace}

\newcommand{\code}[1]{\texttt{\small #1}}

\newcommand{\tool}{\textsc{iPBT}\xspace}
\newcommand{\argmin}{\mathop{\mathrm{arg\,min}}}

\begin{document}

\title{From Natural Language to Executable Properties for Property-based Testing of Mobile Apps
}
\author{Yiheng Xiong}
\affiliation{%
	\institution{ East China Normal University}
	\city{}
	\country{China}}
\email{xyh@stu.ecnu.edu.cn}

\author{Ting Su}
\affiliation{%
	\institution{ East China Normal University}
	\city{}
	\country{China}}
\email{tsu@sei.ecnu.edu.cn}

\author{Jingling Sun}
\affiliation{%
	\institution{University of Electronic Science and Technology of China}
	\city{}
	\country{China}}
\email{jingling.sun910@gmail.com}

\author{Jue Wang}
\affiliation{%
	\institution{Nanjing University}
	\city{}
	\country{China}}
\email{juewang591@gmail.com}

\author{Qin Li}
\affiliation{%
	\institution{East China Normal University}
	\city{}
	\country{China}}
\email{qli@sei.ecnu.edu.cn}

\author{Geguang Pu}
\affiliation{%
	\institution{East China Normal University }
	\city{}
	\country{China}}
\email{ggpu@sei.ecnu.edu.cn}

\author{Zhendong Su}
\affiliation{%
	\institution{ETH Zurich}
	\city{}
	\country{Switzerland}}
\email{zhendong.su@inf.ethz.ch}


\begin{abstract}

Property-based testing (PBT) is a popular software testing methodology and is effective in validating the functionality of mobile applications (apps for short). However, its adoption in practice remains limited, largely due to the manual effort and technical expertise required to specify executable properties. In this paper, we propose a novel structured property synthesis approach that automatically translates property descriptions in natural language into executable properties, and implement it in a tool named \tool.  
Our approach decomposes the problem into UI semantic grounding and executable property synthesis.
It first builds an enriched widget context via multimodal LLMs to align visual elements with their functional semantics, and then uses an LLM with in-context learning to generate framework-specific executable properties.
We evaluate \tool with a closed-source LLM (GPT-4o) and an open-source LLM (DeepSeek-V3) on 124 diverse property descriptions derived from an existing benchmark dataset. \tool achieves 95.2\% (118/124) accuracy on both LLMs. Notably, an ablation study reveals that the enriched widget context contributes to an absolute improvement of up to 20.2\% (from 75.0\% to 95.2\%).
A user study with 10 participants demonstrates that \tool reduces the time required to write executable properties by 56\%, suggesting substantially lower manual effort.
Furthermore, evaluations on 1,180 linguistically diverse variations demonstrate \tool's robustness (87.6\% accuracy), indicating its capability to handle varied expressions.

\end{abstract}
\maketitle
\section{Introduction}

Property-based testing (PBT) has emerged as a powerful testing methodology that validates software program correctness by checking properties.
Unlike example-based testing~\cite{daka2014survey} which relies on specific input-output pairs to determine test outcomes, PBT systematically generates a large number of inputs to verify whether the system under test satisfies the defined properties. 
The pioneering PBT framework QuickCheck~\cite{quickcheck} has inspired many other PBT frameworks that successfully uncover bugs difficult to detect with traditional testing techniques across various software domains.~\cite{arts2006testing,karlsson2020quickrest,hughes2016experiences,hughes2016mysteries,2022quickstrom,santos2018property,kea}.

\begin{figure*}[t]
    \centering
    \includegraphics[width=\textwidth]{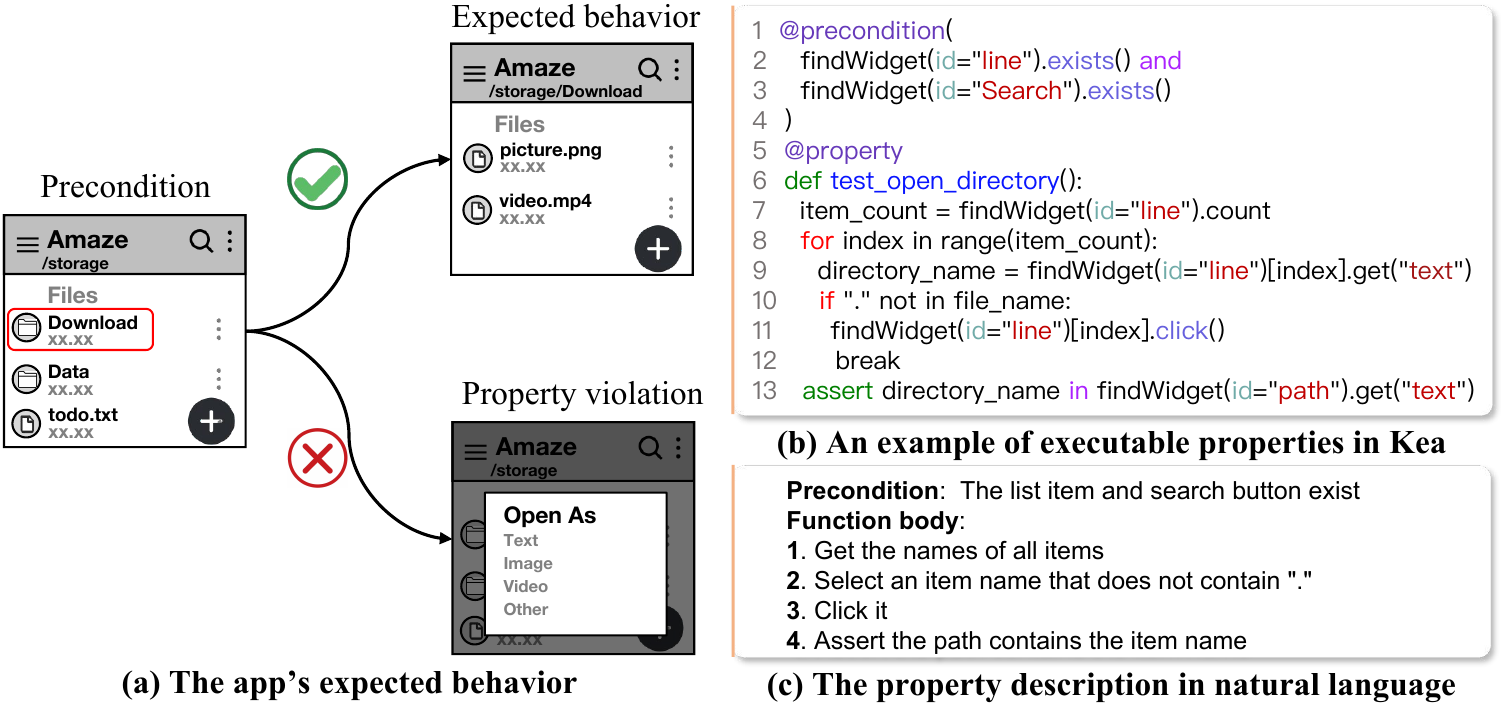} 
    \caption{An example of the property in app Amaze.} 
    \label{fig:property_code_description_example} 
\end{figure*}

Recently, several research efforts have applied PBT in testing mobile apps~\cite{kea,pbfdroid,sun2024property,lam2017chimpcheck} to address the oracle problem, determining whether an app's behavior aligns with expected outcomes. In these work, users specify expected behaviors as executable properties, and then the PBT framework automatically generates GUI event sequences to validate them.
For example, consider Amaze~\cite{AmazeFileManager}, a popular file management app. A typical property is that when a user clicks on a directory (\eg, "Download"), the app should open that directory and display its contents, rather than triggering a file-opening dialog (which is only expected for files). Fig.~\ref{fig:property_code_description_example}(a) illustrates the expected app behavior. Fig.~\ref{fig:property_code_description_example}(b) shows the corresponding executable property written in Kea~\cite{kea}, a recent effective PBT framework for finding functional bugs in mobile apps. The property consists of a precondition (lines 1-4) to check the presence of the file and search button; an interaction scenario (lines 7-12) simulates a click action to open the directory; and a postcondition (line 13) checks whether the path contains the directory name. Then, the PBT framework can generate a large number of GUI events to verify this property and report bugs when the property is violated.

However, these PBT frameworks see limited adoption because testers must translate high-level intents into framework-constrained, executable properties, which demands substantial manual effort and expertise. Testers must learn framework-specific DSLs, including specialized syntax, APIs, and conventions. This imposes a steep learning curve, particularly for those without a strong programming background. 
Moreover, specifying executable properties requires the tedious manual inspection of the app’s view hierarchy to locate low-level UI widget identifiers (\eg, \code{id="Search"}). It also requires correctly implementing complete executable properties under the constraints of the testing framework, both of which are non-trivial and error-prone in practice. Together, these barriers create a significant gap that restricts the widespread application of PBT.

To bridge this gap, we enable testers to specify properties in natural language, which is more intuitive and lightweight. To reduce ambiguity while preserving accessibility, we structure property descriptions in a widely-used Hoare logic format, like Given-When-Then used in Gherkin~\cite{gherkin} for Behavior-Driven Development~\cite{smart2023bdd}. Fig.~\ref{fig:property_code_description_example}(c) illustrates such a structured description. While translating natural language into executable code has been studied, classic rule-based approaches~\cite{thummalapenta2012automating,pandita2012inferring,das2019mynlidb} are insufficient for mobile PBT due to two main limitations. First, rule-based approaches lack flexibility. Enumerating rules to cover diverse paraphrases of user intent (\eg, “open the folder” vs. “navigate into the directory”) does not scale and itself incurs substantial manual effort. Second, these approaches lack semantic grounding. They struggle to accurately map high-level widget descriptions (\eg, “search bar”) to the low-level widget identifiers. 


More fundamentally, these limitations expose a deeper challenge: grounding the informal test intent into concrete executable properties. To address this challenge, we propose a structured property synthesis approach that automatically translates natural-language property descriptions into executable properties.
We decompose this problem into two distinct phases: UI semantic grounding and executable property synthesis.
We first tackle UI semantic grounding by extracting comprehensive GUI information (\eg, view hierarchies and screenshots) from the app and constructing enriched widget contexts for each UI widget. We employ Multimodal Large Language Models (MLLMs) to generate semantic annotations for each widget by jointly reasoning over visual appearance and structural information.
These annotations serve as grounding signals, enabling accurate mapping between user-described widgets (\eg, "item name") and concrete widget identifiers (\eg, "line"), even when the identifiers themselves are not descriptive.
Building on this grounded widget context, we then perform executable property synthesis. We leverage LLMs as inference engines to synthesize executable properties that integrate (i) the property description, (ii) enriched widget context, (iii) framework APIs, and (iv) few-shot demonstrations~\cite{peng2023instruction}. This design eliminates the need for handcrafted rules while enabling the model to adapt to the specific constraints of property generation.

We implemented our approach as a tool named \tool. To evaluate its effectiveness, we conduct experiments with a closed-source LLM (GPT-4o~\cite{gpt4o}) and an open-source LLM (DeepSeek-V3~\cite{deepseek}) on 124 diverse properties from the Kea benchmark~\cite{kea}. \tool correctly synthesizes executable properties for 118 out of 124 cases (95.2\%) with both models.
Notably, our ablation study reveals that the enriched widget context contributed to a 20.2\% absolute increase in accuracy, highlighting its critical role in grounding UI semantics.
Beyond accuracy, a user study demonstrated the practical utility of \tool, reducing property authoring time by 56\% compared to manual composition.
We further evaluate robustness by using another LLM, Llama-3.1~\cite{llama}, to generate 1,180 diverse paraphrased variants of the original property descriptions.
Under these variations, \tool achieves 87.6\% accuracy (1,034/1,180) with GPT-4o and 87.5\% (1,032/1,180) with DeepSeek-V3. These results indicate that \tool is robust to the variability of natural language property descriptions.

In summary, this paper has made the following contributions:
\begin{itemize}[label=\textbullet,leftmargin=*]
    \item At the conceptual level, we introduce a novel approach that reduces the manual effort and lowers the technical barrier for property-based testing of mobile apps.
    \item At the technical level, we have implemented our idea as a tool named \tool that (i) constructs enriched widget contexts via UI semantic grounding to align user-described widgets with concrete UI elements, and (ii) leverages LLMs with in-context learning to synthesize framework-specific executable properties.
    \item At the empirical level, we construct a new evaluation dataset with 
\emph{124 human-written natural language property descriptions derived from real-world bugs}.
We further generate \emph{1,180 linguistically diverse LLM-based variants} for robustness evaluation.
Based on this dataset, we conduct comprehensive evaluations and summarize practical lessons learned on applying LLMs to property-based testing. 

\end{itemize}

\section{Background}

\subsection{Large Language Models}
Large language models (LLMs) have been shown to perform well on a wide range of tasks in natural language processing~\cite{min2023recent} and software engineering~\cite{zhang2023survey,hou2024large}.
LLMs, such as GPT-4o, DeepSeek-V3, are deep neural networks trained on massive amounts of text data, enabling them to generate human-like text, understand complex queries, and perform different tasks. These models are typically based on the Transformer architecture~\cite{vaswani2017attention}, which relies on self-attention mechanisms to process and generate text efficiently at scale.
Recently, Multimodal Large Language Models (MLLMs) have emerged, which extend the capabilities of traditional LLMs by incorporating additional modalities such as images, audio, or structured data alongside text~\cite{mllm_survey}. This multimodal integration allows MLLMs not only to process natural language but also to understand and reason about visual or other non-textual inputs. In the context of GUI testing, MLLMs are particularly useful as they can jointly leverage textual descriptions and GUI screenshots to better interpret UI semantics.

\subsection{In-Context Learning}
LLMs are typically pre-trained on large corpora of text and code data. To adapt LLMs on customized tasks, fine-tuning~\cite{devlin2019bertpretrainingdeepbidirectional}, which requires training on a pre-trained model with additional massive data,  and prompt engineering~\cite{peng2023instruction} (\eg, chain-of-thought~\cite{wei2022chain}, in-context learning~\cite{peng2023instruction}, and multi-step reasoning~\cite{zhou2022least} ) are two common approaches. 
In-context learning refers to the ability of LLMs to perform customized tasks with a few examples provided in the input prompt, without requiring model training.
In our approach, we adopt in-context learning, which enables LLMs to perform custom tasks using just a few examples included directly in the input prompt.  



\begin{figure}[t]
    \centering
    \includegraphics[width=\textwidth]{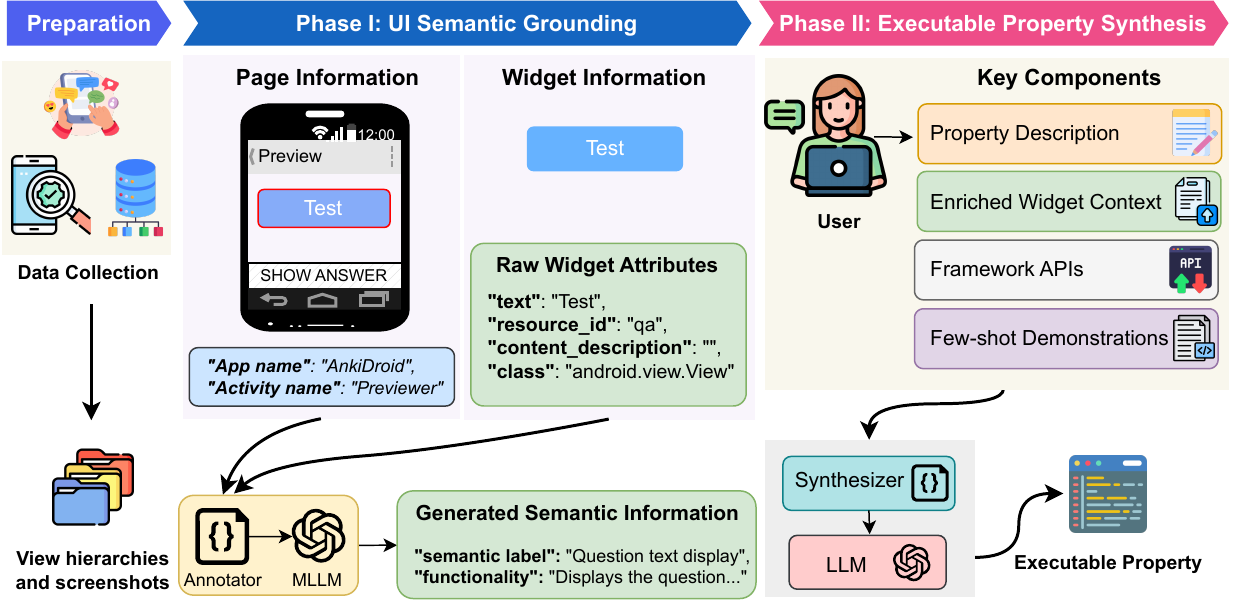} 
    \caption{Overview of \tool.} 
    \label{fig:overview} 
\end{figure}

\section{Approach}

At a high level, \tool operates as a structured synthesis approach designed to translate natural language property descriptions into executable properties. 
Fig.~\ref{fig:overview} presents the overall workflow of \tool, which consists of two main phases: (i) the \textit{UI Semantic Grounding} phase (\S\ref{sec:ui_widget_context_construction}), which extracts the GUI information of the app under test, leverages MLLMs to generate functionality annotations for widgets, and constructs the enriched widget context for each UI element; and (ii) the \textit{Executable Property Synthesis} phase (\S\ref{sec:executable_property_generation}) that generates executable properties by encoding user APIs, UI widget identifiers, property description and examples into a carefully designed prompt. Collectively, these two phases enable \tool to effectively bridge the gap between informal user intent and concrete implementation details for property-based testing. We next describe each phase in detail.

\subsection{Phase I: UI Semantic Grounding}
\label{sec:ui_widget_context_construction}
In Android app testing, UI widget identifiers (\eg, resource\_id, text, and content\_description) are commonly used to locate and interact with target widgets displayed on the screen.
Consequently, a critical prerequisite for synthesizing executable properties is grounding high-level natural language descriptions onto these concrete identifiers.
For instance, an instruction like ``click the Settings button'' must be accurately mapped to a specific widget, such as \code{text="Settings"} or \code{id="action\_settings"}. However, relying solely on raw identifiers is often insufficient. In practice, these identifiers are frequently opaque, poorly maintained, or generic (\eg, \code{id="button1"}), failing to reflect the widget's true functionality.
To bridge this semantic gap, we extract comprehensive GUI information and leverage Multimodal LLMs (MLLMs) to generate semantic functionality annotations for each widget.
The resulting enriched widget context consists of raw attributes (\eg, \code{text}, \code{resource\_id}) with MLLM-inferred semantics. This context serves as a robust bridge, enabling the subsequent executable property synthesis phase to accurately match user-described widgets to the correct identifiers.


\subsubsection{Widget Context Extraction}
\label{sec:widget_context_extraction}
In Android apps, the layout defines the structure of each page, while widgets (\eg, \code{Button}) handle user-triggered events (\eg, \code{onClick})~\cite{android_layout_widget}. To extract the contextual information of these widgets, the first step is to collect the necessary GUI data (view hierarchies and screenshots). Notably, our approach supports GUI data acquisition via various methods, including: (1) manual interaction by testers; (2) automated exploration using testing tools; and (3) direct provision from app vendors.
From the collected GUI data, we extract two complementary components that facilitate MLLM understanding and the generation of functionality annotations for widgets:

\begin{itemize}[label=\textbullet,leftmargin=*]
\item \textbf{Page information} provides high-level context about the app page under test, including the \emph{app name}, \emph{activity name}, and \emph{page screenshot}. Specifically, the \emph{app name} typically reflects the domain or type of the app (\eg, SimpleNote implies a note-taking app), offering prior knowledge about its overall functionality. The \emph{app name} is obtained by statically analyzing the  \code{AndroidManifest.xml} file. The \emph{activity name}, parsed from the runtime layout file, specifies the functionality of the current page within the app (\eg, \code{LoginActivity} for user authentication, \code{SettingsActivity} for configuration). The \emph{page screenshot}, captured via the Android Debug Bridge (ADB), preserves the visual layout and arrangement of the UI widgets. It serves as a direct reference that complements textual and structural information. The target widget is highlighted with a red bounding box in the screenshot to help the MLLM accurately locate it. 

\item \textbf{Widget information} provides fine-grained details about the interactive widgets on each page. It consists of two parts: the cropped widget image and the widget attributes.  The \emph{widget image} is obtained by cropping the page screenshot according to the widget bounds. The \emph{widget attributes} are extracted from the view hierarchy file. We focus on four commonly used fields: "\code{text}", "\code{resource\_id}", "\code{content\_description}", and"\code{class}", as they provide valuable signals about the intended behavior of the widget. Specifically, the "\code{text}" field corresponds to the string displayed on the widget (\eg, "Login"), often directly indicating its functionality. The "\code{resource\_id}" is a developer-assigned identifier that may encode semantic hints about the widget’s role or logical grouping (\eg, \code{id=btn\_submit}). The "\code{content\_description}" field, typically designed for accessibility~\cite{android_accessible}. Finally, the "\code{class}" specifies the Android class type of the widget (\eg, \code{Button}, \code{EditText}, \code{ImageView}), which reflects the type of interaction it supports. 

\end{itemize}

These two components together provide complementary structural and visual information, enabling the MLLM to better reason about widget semantics. Crucially, the extraction of these artifacts is fully automated. The extracted widget context serves as the foundation for the subsequent phase, where MLLMs are leveraged to generate accurate functionality annotations for each widget.

\begin{table}[t]
	\caption{The prompt design for generating functionality annotation of the widget.}
	\centering
	\begin{adjustbox}{max width=\textwidth}
		
\begin{tabular}{m{1em}|m{4cm}|m{12cm}}
\toprule
\textbf{ID} & \textbf{Prompt Component}           & \multicolumn{1}{c}{\textbf{Instantiation}}                                                                                                                                                                         \\ \midrule
\ding{172}   &  Role  Assignment                & You are a professional mobile app UI semantic annotation assistant.  \\ \midrule
\ding{173}  & Task & Please annotate the provided UI
widget with the semantic label and functionality description based on the given context. \newline - The full page screenshot, where the target widget is highlighted with a red box.
\newline - The cropped widget image and its attributes.
\newline - The provided app name and foreground activity name.
\\ \midrule
\ding{174}  & Few-shot Demonstrations & [example input] + [example output]  \\ \midrule 

\ding{175}  & Input & [page information]+[widget information]
\\ \midrule

\ding{176}  & Constraints & [Strict rules]
                                \\  \bottomrule
\end{tabular}
	\end{adjustbox}
	\label{fig:prompt_widget}
\end{table}


\subsubsection{Widget Functionality Annotations Generation}
Leveraging the extracted widget context, we employ an MLLM to synthesize functionality annotations. 
We construct a structured prompt to guide the model, as illustrated in Table~\ref{fig:prompt_widget}.
The prompt is organized into five components: Component \ding{172} defines the role of the model to establish a professional persona. Component \ding{173} specifies the Task, directing the MLLM to infer semantics based on the provided visual and structural context.
Component \ding{174} provides two demonstrations that serve as the concrete reference selected from the SimpleNote app~\cite{simplenote}. They illustrate the expected mapping from the raw widget context to the target semantic label and functionality description. Empirically, we found that two representative demonstrations are sufficient for the MLLM to grasp the task requirements and generate high-quality annotations. 
Component \ding{175} gives the input of context, including page information and widget information extracted in the previous step. Component \ding{176} outlines the constraints to ensure output consistency.

Given these instructions, the MLLM generates two key fields for each widget: \emph{semantic label} and \emph{functionality}. The \emph{semantic label} is a concise identification of the widget, \eg, login button, or settings option. The \emph{functionality} offers a description of its behavior, \eg, allows the user to log into their account, or navigates to settings screen. 

For example, consider the target widget shown in the green box in Fig.~\ref{fig:overview}. Sole reliance on these raw attributes is insufficient to deduce its functionality. To resolve this ambiguity, we construct an enriched widget context by aggregating page information, such as the app name \texttt{AnkiDroid} and activity \texttt{Previewer}, with the widget information. This combined information enables the MLLM to effectively anchor the widget's semantics. Consequently, \tool produces functionality annotations, augmenting the original attributes with a semantic label ("Question text display") and a precise functionality description ("Displays the question text to the user for review"). 
These enriched widget contexts serve as the foundation for generating executable properties in the next phase.

\subsection{Phase II: Executable Property Synthesis}
\label{sec:executable_property_generation}
\subsubsection{Writing Property Descriptions}
To ensure that property descriptions are both intuitive for testers and structured for synthesis, we adopt a Hoare logic-style representation. This style aligns with industry-standard practices, such as Gherkin~\cite{gherkin} widely employed in Behavior-Driven Development (BDD)~\cite{smart2023bdd}.
In the context of PBT in mobile apps, an executable property is formalized as a triple $\langle P, I, Q \rangle$, where (1) $P$ is the precondition, which defines the when could check the property, (2) $I$ is the interaction scenario, which defines the sequence of user actions to execute the target functionality, and (3) $Q$ is the postcondition, which specifies the expected UI state after the interaction. 
To facilitate this formalization, we structure the natural language property description into two segments:
\begin{itemize}[label=\textbullet,leftmargin=*]
\item \textbf{Precondition:}
This segment maps directly to $P$. It describes the initial visible state required to trigger the property (\eg, the file name exists'').
\item \textbf{Function Body:} This segment encapsulates both $I$ and $Q$. It contains the execution steps of the target functionality (\eg, select a file name that does not contain `.' and click it'') and explicitly states the expected effects (\eg, ``verify the path contains the file name'').
\end{itemize}


\begin{table}[t]
	\caption{The prompt design for generating executable properties.}
	\centering
	\begin{adjustbox}{max width=\textwidth}
		
\begin{tabular}{m{1em}|m{4cm}|m{12cm}}
\toprule
\textbf{ID} & \textbf{Prompt Component}           & \multicolumn{1}{c}{\textbf{Instantiation}}                                                                                                                                                                         \\ \midrule
\ding{172}  &  Role Assignment                 & You are an expert in Python programming and Android app testing, and your role is to write test snippets for Android apps.                                                                                                      \\ \midrule
\ding{173}  & Framework APIs             & \textbf{The following APIs are available for writing property:} \newline widget: findWidget(identifier); click: widget.click()  long click: widget.long\_click(); get text: widget.get("text"); exists: widget.exists(); back: press("back"); ...                                                                                        \\ \midrule
\ding{174}  & Enriched Widget Context & \textbf{The app's UI widget identifiers are detailed below for reference, ensuring accurate element selection in tests}:\newline \{    "text": "Download",       "resource\_id":"line",     "description": "null",     "class": "android.widget.TextView",    "semantic label": "File name text",    "functionality": "Display the name of the file" \}, ...\\
\midrule
\ding{175}  & Few-shot Demonstrations & Here are the two example test snippets that you might write, based on the given property descriptions: {[}Example property description and executable properties{]}                                                  \\ \midrule
\ding{176}  & Property Description  & \textbf{Your task: Using the available APIs, UI widget identifiers and following the example format, please write a test snippet for the following property:}\newline \textbf{Precondition}: The list item and search button exist \newline \textbf{Function body:} \newline 1. Get the names of all items \newline 2. Select an item name that does not contain "."\newline 3. Click it\newline 4. Assert the path contains the item name \\ \midrule
\ding{177}  & Constraints           &  Respond only with the Python code, strictly adhering to the given property description. Do not include any explanations, comments, or text outside the code block.                                                     \\  \bottomrule
\end{tabular}
	\end{adjustbox}
	\label{fig:prompt1_design}
\end{table}


\subsubsection{Constructing Prompt and Generating Executable Properties.}

Table~\ref{fig:prompt1_design} illustrates the structured prompt employed by \tool to synthesize executable properties. The prompt comprises six distinct components, labeled \ding{172} through \ding{177}, designed to guide the LLM's synthesis process:

\begin{itemize}[label=\textbullet,leftmargin=*] \item \textbf{Role Assignment (\ding{172}):} This component defines a specialized persona for the LLM, priming it to focus on the domain of mobile app testing and code generation.
\item \textbf{Framework APIs (\ding{173}):} To ensure the synthesized code is syntactically valid, we provide the complete list of user-facing APIs supported by the target framework (Kea in our implementation). We curated the set of APIs like \code{click()} and \code{exists()} by analyzing the framework's official documentation and source code. Crucially, these APIs are **app-agnostic**, allowing this component to be reused across different apps. Note that this component is modular and can be substituted with APIs from other PBT frameworks if desired.

\item \textbf{Enriched Widget Context (\ding{174}):} This component supplies the Enriched Widget Context constructed in Phase I. By integrating raw identifiers with MLLM-generated semantic annotations, this section enables the LLM to ground the natural language descriptions to the correct UI widget identifiers.

\item \textbf{Few-shot Demonstrations (\ding{175}):} To facilitate in-context learning, we crafted two concrete examples derived from the SimpleNote app~\cite{simplenote}. Our selection follows two principles to ensure robustness: (1) To avoid bias, the examples originate from an app distinct from the subject apps used in our evaluation. (2) The examples representatively demonstrate the usage of key APIs (\eg, \code{findWidget}, \code{exists}) and the mapping to the target Hoare logic structure ($\langle P, I, Q \rangle$).
Empirically, we found that two diverse examples are sufficient for the model to generalize the generation pattern while maintaining token efficiency.

\item \textbf{Property Description (\ding{176}):} The component describes the task with the property description and highlights that the output should follow the format of the example executable properties.

\item \textbf{Constraints (\ding{177}):} To ensure the output is machine-readable, we enforce strict constraints. We explicitly instruct the LLM to generate \textit{only} the code snippet without verbose explanations or markdown formatting, which facilitates the automated extraction of the generated properties.
\end{itemize}

Upon construction, the complete prompt is fed into the LLM, which then synthesizes the final executable properties.

\section{Evaluation}
We aim to answer the following research questions:

\begin{itemize}[label=\textbullet,leftmargin=*]
    \item \textbf{RQ1}: What is the correctness of executable properties generated by \tool based on property descriptions? How important are the widget functionality annotations in affecting the correctness?

    \item \textbf{RQ2}: To what extent can \tool reduce manual effort? What are the differences in the complexity of the natural language property descriptions and executable properties generated by \tool?
    
    \item \textbf{RQ3}: How robust is \tool in generating executable properties based on diverse property descriptions?
    
\end{itemize}

\noindent{\textbf{Large Language Models.}}
We evaluate our approach using two representative large language models for executable property generation: one closed-source model (GPT-4o) and one open-source model (DeepSeek-V3). This selection allows us to assess the effectiveness of our approach across different model ecosystems. In addition, we employ a multimodal large language model (GPT-4o mini) to generate functionality annotations for UI widgets, as it supports both textual and visual inputs.

\noindent{\textbf{Writing property descriptions.}}
To evaluate our approach, we construct property descriptions based on the dataset from the prior work Kea~\cite{kea}, which provides 124 diverse executable properties across eight popular open-source Android apps covering diverse app categories, \eg, tools, editor, education, and audio player. 
Importantly, each property is derived from a distinct real-world historical bug, ensuring that the dataset reflects practical issues encountered in diverse app contexts. Since these bugs span different functional modules of the apps (\eg, navigation, data management, and configuration), the resulting properties capture a broad spectrum of app behaviors. Moreover, as a benchmark originally curated for evaluating property-based testing of Android apps, this dataset offers both diversity and practical relevance, making it well-suited for our evaluation.
The dataset also includes the corresponding bug reports to facilitate understanding and reproduction of each bug. 

To construct the natural language property descriptions, we followed a three-step process:
(1) Property collection. 
We collected 124 properties from Kea's dataset, including the executable properties, associated bug reports, and APK files.
(2) Property understanding.
Each line of an executable property typically represents a UI event, making it difficult to infer the corresponding widget based solely on its identifier. To gain a comprehensive understanding, we manually installed the associated APK files on mobile devices and interacted with the apps to reach the states satisfying the precondition. Then, we executed the executable properties on the app to observe each step. This process can help us understand each component of the executable properties, including the precondition, interaction scenario, and postcondition.
(3) Property description construction.
One author wrote each property description in natural language based on the executable properties and observed app behavior. To ensure clarity and correctness, the remaining co-authors discussed together to resolve any inconsistencies or ambiguities through iterative revision.

\noindent{\textbf{Constructing enriched widget context.}} The first step of \tool aims to construct the enriched context for the widgets in each app.
While our approach supports diverse data collection methods (as detailed in Section~\ref{sec:widget_context_extraction}), we employed an automated strategy in our evaluation to minimize manual intervention. We utilized DroidBot~\cite{wen2023droidbot}, a popular open-source automated GUI testing framework, extended with a random exploration strategy, to perform a three-hour exploration on each app~\cite{su_stoat_2017}.
We acknowledge that achieving full coverage via automated exploration remains a long-standing challenge in both research and practice. However, advancing exploration algorithms is orthogonal to our primary contribution of property generation. To decouple the effectiveness of our generation approach from the limitations of the exploration tool, we additionally executed the "main path" provided in the dataset, comprising events from the app entry to the state satisfying the property's precondition. This ensures that all relevant UI widgets involved in the target properties are captured, providing a fair basis for evaluating \tool's generation capabilities. Finally, \tool extracts the page and widget information and leverages GPT-4o mini to construct the enriched context.


\noindent{\textbf{Correctness of the generated executable properties.}}
The 124 executable properties in the Kea dataset are derived from 124 historical bugs. Thus, verifying whether the generated executable properties by \tool can reproduce the associated bug serves as a primary indicator of its correctness. Specifically, for each executable property, we follow these steps: (1) Start from the app’s entry point and navigate along the bug-triggering path.
(2) Interact with the app until reaching the state that satisfies the preconditions described in the executable property. 
(3) Execute the generated executable properties to determine whether it successfully triggers the historical bug.
The whole process can be automatically performed by running the scripts (from app entry to the state satisfying the precondition) and executable properties in Kea. 

For a few cases, even the generated executable properties are able to trigger the historical bug, they may not faithfully capture the intended logic of the original property. For example, the generated executable property may omit some conditions in the precondition. To address this, we additionally evaluate whether the generated properties preserve original intent from the following two dimensions:
\begin{itemize}[leftmargin=*]
    \item \textbf{Precondition and postcondition.} We verify whether the clauses and logical operators in the generated executable properties match those in the ground-truth. Missing or incorrect preconditions or postconditions can lead to inconsistencies in specific scenarios.
    \item \textbf{Interaction scenario.} First, we check whether the event sequence in the interaction scenario matches the ground-truth, where each event contains the event action and the target UI widget. Second, for executable properties containing conditional branches, we check whether these branches are consistent with the intended logic. Extra or missing branches may still allow bug reproduction but fail to reflect the precise execution flow of the original executable property.
\end{itemize}

To ensure reliability, two co-authors of this paper independently evaluated each executable property. 
We measured inter-rater agreement using Cohen’s Kappa, achieving a substantial agreement ($\kappa = 0.96$). 
Any disagreements were resolved through discussion.

In Android app testing, UI widgets are located using identifiers (\eg, \code{resource\_id}, \code{text}). Therefore, the same UI widget can be matched through different identifiers. For example, \code{text="Settings"} and \code{resourceId="app.settings"} both refer to the same button, which navigates to the system configuration interface when clicked. When evaluating the widget matching in the generated executable properties, we treat such cases as correct matches as long as the different attributes refer to the same underlying widget.

\subsection{RQ1: Correctness}
\noindent{\textbf{Evaluation setup.}}
To evaluate the correctness of \tool in generating executable properties, we constructed 124 prompts with 124 property descriptions based on the designed prompt template (Table~\ref{fig:prompt1_design}). 
To ensure deterministic and stable results, we set the temperature parameter of the LLM to 0, a setting commonly used in prior work~\cite{yang2024evaluation,fan2023large,ouyang2023llm}. For each prompt, we sequentially invoked the LLM and recorded the generated executable properties as the output. 
Moreover, we conduct an ablation study to evaluate the contribution of the widget functionality annotation module.
Specifically, we removed the functionality annotation of each widget and then repeated the same experimental procedure to assess its impact. We also conduct the ablation study in RQ2 and RQ3 in the same process.

\begin{figure}[t]
    \centering
    \includegraphics[width=\textwidth]{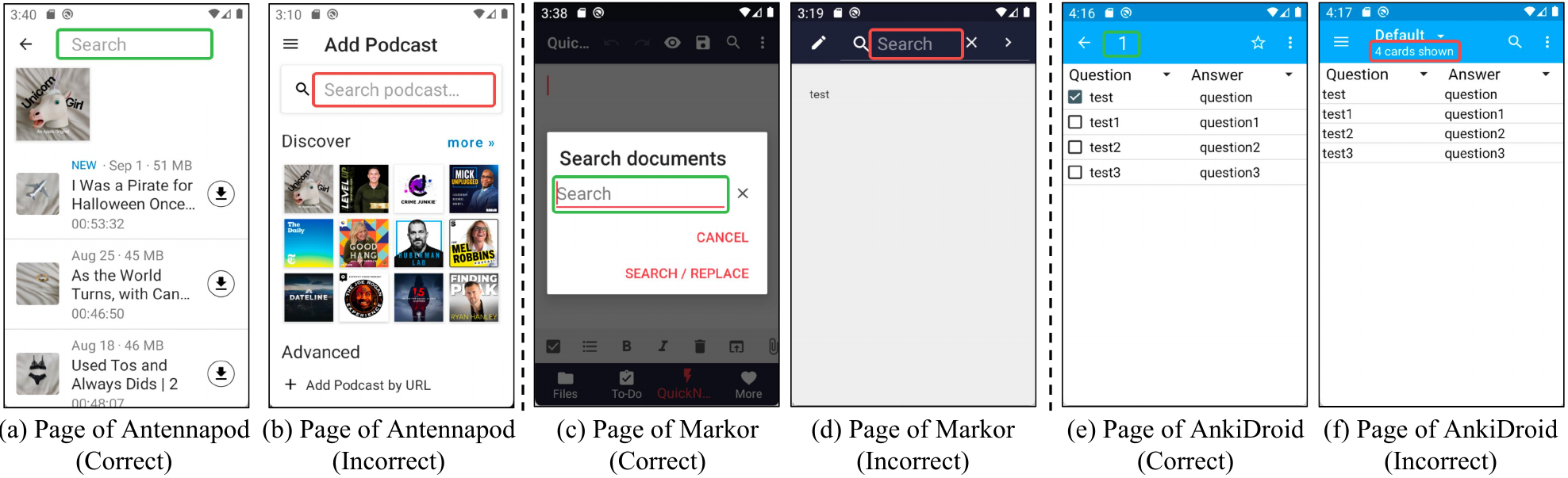} 
    \caption{Examples of failure cases. The green boxes annotate the correct widgets and red boxes annotate the incorrect widgets.} 
    \label{fig:rq1_failure} 
\end{figure}

\noindent{\textbf{Evaluation results.}}
Based on our experimental statistics, for all 124 property descriptions, both GPT-4o and DeepSeek-V3 successfully generated 118 correct executable properties, achieving an accuracy rate of 95.2\%. These results demonstrate the effectiveness of \tool and highlight its ability to reliably translate natural language descriptions into executable properties. 

For the six failure cases, all errors manifested as incorrect widget identifiers generated by \tool. Upon further analysis, we found that in five cases, the failures were caused by the presence of other widgets within the same app that shared similar functionality with the target widget, which misled \tool during the matching process. In the remaining case, the error originated from an ambiguous functionality annotation generated by \tool, which subsequently caused the executable property to be matched to the wrong widget. Fig.~\ref{fig:rq1_failure} presents three examples from different apps, where the correct widgets are highlighted with green boxes and incorrect ones with red boxes. The first two cases (Fig.~\ref{fig:rq1_failure}(a)--(d)) failed because different widgets in the same app shared similar functionality. The third case (Fig.~\ref{fig:rq1_failure}(e)--(f)) failed due to ambiguity in the functionality annotation generated by \tool. Specifically, \tool produced the semantic label \textit{"Current card number"} and the functionality description \textit{"Indicates the number of the current card being viewed"}. In reality, the widget represents the number of the currently selected card, which differs subtly from the generated annotation and led to an incorrect mapping.

\textit{Ablation study.}
The result shows that after removing the functionality annotation of widgets, \tool successfully generated 93 and 96 correct executable properties on GPT-4o and DeekSeek-V3, corresponding to accuracy rates of 75.0\% and 77.4\%, respectively. Compared with the full setting (118 correct executable properties, 95.2\% accuracy), this represents a performance drop of 20.2 and 17.8 percentage points, respectively. This substantial decrease highlights the importance of functionality annotations in helping the LLM accurately understand widgets and generate correct executable properties.

\subsection{RQ2: User study}

\noindent{\textbf{Evaluation setup.}}
In RQ2, we conduct a controlled human-subject experiment to assess how \tool actually supports users in practice.
Specifically, participants were asked to write executable properties manually, as well as to provide property descriptions for \tool to generate executable properties.
We then analyzed their time cost and correctness across various tasks to evaluate how \tool can save manual effort. Our experimental design follows prior work~\cite{yang2021subtle,liang2021interactive,pbfdroid}.

\textit{Dataset of the user study.} We selected 124 executable properties from RQ1 and categorized them into three groups (low, medium, and high complexity) using the metric defined in Kea. This metric accounts for the number of logical clauses and operators in the pre- and postconditions, as well as the number of events in the interaction scenario.
From each complexity group, we randomly selected 2 executable properties from different apps, resulting in a final dataset of 6 executable properties: Property 1 and Property 2 (low complexity), Property 3 and Property 4 (medium complexity), and Property 5 and Property 6 (high complexity). By selecting executable properties from different complexity levels, we ensure that the evaluation captures diverse levels of difficulty while keeping the number of tasks manageable for participants. 
For each property, we installed the corresponding app on an Android emulator and recorded a video walkthrough. Each video demonstrated how to: (1) reach the state that satisfies the precondition, (2) perform the functionality in the interaction scenario, and (3) verify the expected behavior in the postcondition.

\begin{table}[t]
	\caption{Groups in user study}
	\centering
	\begin{adjustbox}{max width=\textwidth}
		\begin{tabular}{lcc}
			\toprule
			\textbf{Group (Participant ID)}         & \textbf{Writing property descriptions}  & \textbf{Writing executable properties}  \\ \midrule
			
			\textsc{Group A (P1-P5)}        & Property 1, 3, 5  & Property 2, 4, 6  \\    

                \textsc{Group B (P6-P10)}        & Property 2, 4, 6  & Property 1, 3, 5 \\     
               \bottomrule
		\end{tabular}
	\end{adjustbox}
	\label{table:groups}
\end{table}

\textit{Participants.} We recruited 10 participants for our user study, a sample size similar to previous related work~\cite{pbfdroid,chen2018ui,zhao2019recdroid} (which recruited 10, 8, and 12 participants, respectively). All participants are graduate students majoring in software engineering, with at least four years of programming experience and familiarity with Python programming. This ensured that they had the necessary technical background to understand and write the executable properties without using \tool. A prior study has shown that graduate students can serve as professionals in software engineering tasks~\cite{salman2015students}. In addition, none of the participants was from the authors of this paper.

\textit{Procedure.} We designed the study as a conventional within-subject controlled experiment, in which each participant was required to write both property descriptions and executable properties. To avoid learning bias caused by increased familiarity, no participant wrote both the description and executable version of the same property. To achieve this, the 10 participants were evenly divided into two groups (Group A and Group B) with comparable programming expertise, and each group was assigned different tasks for the same property (as shown in Table~\ref{table:groups}). In total, this resulted in 60 property-writing tasks.


At the beginning, participants attended a dedicated tutorial introducing the study’s background, the concept of properties in Android apps, and the procedures for writing them. Following the tutorial, each participant was given an example property video and asked to write both the property description and the corresponding executable properties. This warm-up exercise, which lasted approximately 45 minutes, familiarized participants with the task and provided step-by-step guidance.

After the tutorial, participants independently completed six distinct tasks. For each task, they were provided with recorded videos containing the necessary information. 
The study was conducted in a preconfigured desktop environment, where participants used Visual Studio Code to write their property descriptions and executable properties. To ensure fairness, code auto-completion features were disabled.

We recorded the time each participant spent on the tasks (including the time spent on writing property descriptions or executable properties, reading the documentation when needed, and self-checking the written properties). We also collected all written property descriptions and executable properties.
For each collected property description, we leverage \tool to generate the corresponding executable properties. The correctness of the generated code was then assessed using the evaluation metrics described in our experimental setup.

\textit{Complexity of property description and code.} We measured the complexity of the 124 property descriptions and their corresponding executable properties using character count. For instance, the natural language step "click the undo button" has a complexity of 21 characters. This measurement allows us to quantitatively compare the writing effort required for natural language descriptions versus executable properties.

\begin{figure}[t]
    \centering
    \includegraphics[width=0.5\textwidth]{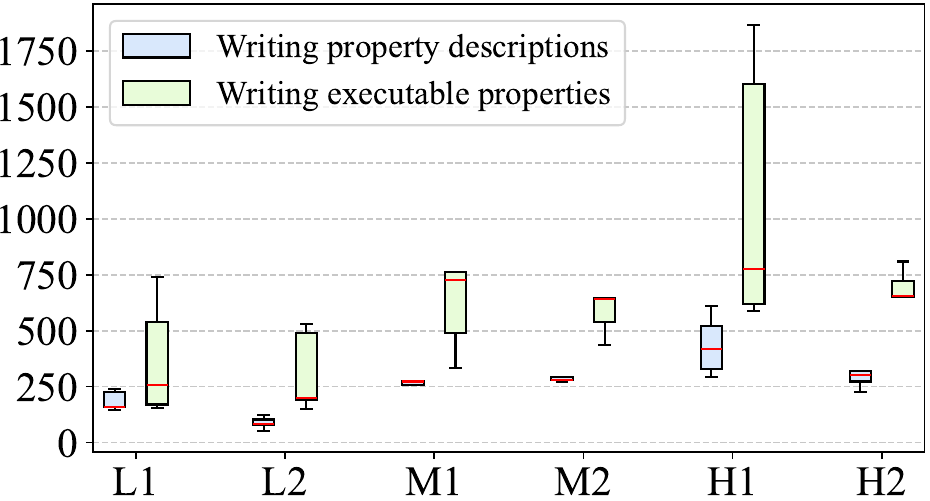} 
    \caption{Time consumption of participants in writing property descriptions and executable properties. The x-axis represents different properties, and the y-axis represents the time spent (in seconds). } 
    \label{fig:time_in_user_study} 
\end{figure}

\noindent{\textbf{Evaluation results.}}
We conducted a detailed analysis of the time efficiency and correctness of executable property generation under two settings: (1) writing property descriptions in natural language and generating executable via \tool, and (2) manually writing executable properties.
\begin{itemize}[label=\textbullet,leftmargin=*]
    \item \textbf{Time efficiency}: Fig.~\ref{fig:time_in_user_study} shows the time spent per task in both approaches. We can see that writing natural language property descriptions significantly reduces the time required to produce executable properties. On average, our approach achieved a 56\% reduction in time, compared to manual writing property descriptions (272.7s vs 625.7s). For low-complexity executable properties (\eg, L2), time savings reached up to 72\%, highlighting the efficiency gains of leveraging LLMs for code generation.
    \item \textbf{Correctness}: Table~\ref{table:accuracy_in_user_study} presents the number of correctly generated executable properties in different approaches. Writing property descriptions with \tool produced 29 correct executable properties, outperforming the manual approach, which yielded 26 correct executable properties. One incorrect executable property generated by \tool approach stemmed from a user typo (a long-click was mistakenly written as a click). In contrast, most errors in the manually written executable properties were caused by participants' unfamiliarity with the Kea framework, such as incorrect API usage in postconditions.
\end{itemize}

\begin{table}[t]
	\caption{The correctness of the executable properties generated by different ways.}
	\centering
	\begin{adjustbox}{max width=\textwidth}
		\begin{tabular}{lccccccccc}
			\toprule
			\textbf{Task}                                           & \textbf{LLM}                       & \textbf{Setting}            & \textbf{L1} & \textbf{L2} & \textbf{M1} & \textbf{M2} & \textbf{H1} & \textbf{H2} & \textbf{Total} \\ \midrule 
\multirow{4}{*}{Writing property descriptions} & \multirow{2}{*}{GPT-4o}   & with annotations    & 5  & 5  & 5  & 4  & 5  & 5  & \textbf{29}    \\
                                               &                           & \textit{without annotations} & 4  & 5  & 5  & 1  & 4  & 5  & 24    \\ \cline{2-10} 
                                               & \multirow{2}{*}{DeepSeek-V3} & with annotations    & 5  & 5  & 5  & 4  & 5  & 5  & \textbf{29}    \\
                                               &                           & \textit{without annotations} & 4  & 5  & 0  & 1  & 4  & 2  & 16    \\ \hline
Writing executable properties                   & \multicolumn{2}{l}{}                           & 5  & 5  & 4  & 5  & 3  & 4  & 26    \\ \hline
			
		\end{tabular}
	\end{adjustbox}
	\label{table:accuracy_in_user_study}
\end{table}

To better understand these differences, we identified two main contributing factors. First, when manually writing executable properties, participants needed additional time to inspect UI attributes for widget identification, while \tool automated this process based on matching widgets from property descriptions. Second, even though all participants received training on Kea, we observed frequent consultation of the documentation during manual writing, especially for framework-specific APIs. By contrast, \tool allows users to focus on specifying intended behaviors in natural language, while delegating low-level implementation details to the LLM. This not only improves efficiency but also reduces errors caused by limited familiarity with the framework.

\textit{Ablation study.} In Table~\ref{table:accuracy_in_user_study}, we can see that without functionality annotations, \tool only generated 24 and 16 correct executable properties, respectively. This result indicates the necessity of the designed functionality annotations in our approach.

\begin{figure*}[t]
    \centering
    \includegraphics[width=0.9\textwidth]{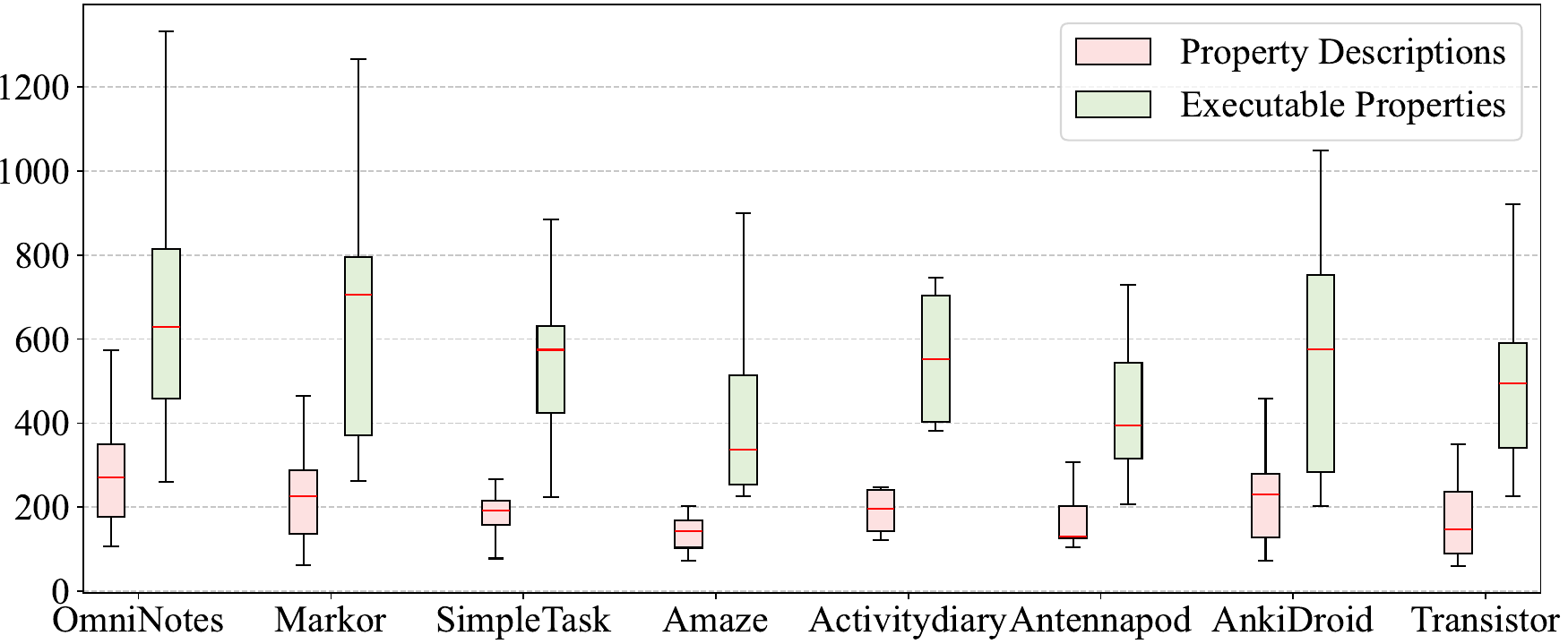} 
    \caption{Comparison of the complexity (measured in character count) between natural language property descriptions and their corresponding executable properties across eight Android apps.} 
    \label{fig:property_description_code_complexity} 
\end{figure*}

\textit{Complexity comparison.} Fig.~\ref{fig:property_description_code_complexity} presents the complexity comparison between the 124 natural language property descriptions and their corresponding executable properties, across eight popular Android apps. The x-axis denotes the app names, while the y-axis represents the complexity, quantified by the number of characters.
On average, the property descriptions contain 211.3 characters, whereas the corresponding executable properties contain 555.0 characters. 
This significant difference in complexity highlights the efficiency of the natural language-based approach. Property descriptions are inherently easier to write than their corresponding executable properties, further emphasizing the practicality of \tool in executable properties generation.

\subsection{RQ3: Robustness}

\noindent{\textbf{Evaluation setup.}}
Building on the user study in RQ2, which evaluates how \tool supports users in practice, RQ3 serves as a complementary experiment that examines robustness. 
While RQ2 considers the property descriptions actually written by participants, RQ3 evaluates whether \tool can generate correct executable properties when faced with semantically equivalent but differently worded property descriptions, simulating the natural variation in how different users might express the same intent.


Prior work has shown that LLMs can generate high-quality paraphrases with greater lexical and syntactic diversity than those produced by crowd workers~\cite{cegin2023chatgpt,berro2025llms}.
Following this line of work, we use an LLM to automatically generate paraphrased versions of each original property description.

To avoid bias from using the same model for both paraphrasing and code generation, we employ a different LLM, Llama-3.1-405B, for paraphrase generation.
For each property description, we invoke Llama-3.1 multiple times to generate a pool of candidate paraphrases.
Specifically, for each property description, we invoke the Llama-3.1 10 times using the designed prompt (the prompt is provided in the artifact due to space limitations), generating 10 paraphrased descriptions per call, resulting in a total of 100 paraphrased variations per property description. 

However, the LLM-generated variations may not be mutually diverse between them. To select a diverse subset of 10 paraphrases per property description, we employed the BLEU score~\cite{papineni2002bleu} to quantify lexical similarity and guide the selection process.
The BLEU score was originally developed for assessing machine translation quality by measuring the similarity between machine-generated translations and human-written reference translations. 
In the context of paraphrasing, a lower BLEU score indicates higher diversity compared to the original text.
Specifically, we adopted a greedy selection strategy to identify a set of paraphrases that are mutually diverse. First, we identified the two most mutually diverse paraphrases (\ie, the pair with the lowest BLEU score between them) to initialize the selection set $S$. Then, in each subsequent iteration, we computed the Self-BLEU score~\cite{zhu2018texygen} between each remaining candidate and the current set $S$, which quantifies similarity to the existing set. The candidate with the lowest Self-BLEU score was added to $S$. This process continued until $S$ contained 10 paraphrased descriptions.
Formally, the objective of our selection process is to identify a subset 
$S$ $\subseteq$ $C$ of size $k$, where $C$ is the set of all candidate paraphrases and $k$ = 10) that minimizes the Self-Bleu score:
\[
\argmin_{S \subseteq C,\ |S| = k} \ \frac{1}{|S|(|S| - 1)} \sum_{\substack{x, y \in S \\ x \ne y}} \text{BLEU}(x, y)
\]
This selection strategy ensures that the chosen paraphrases are not only diverse relative to the original text but also mutually diverse within the set, resulting in a robust dataset for evaluating our approach under varied natural language expressions.

In RQ3, we focus on 118 property descriptions from RQ1 that were successfully translated into correct executable properties. Using the paraphrasing procedure described above, we generate 10 paraphrases for each property description, resulting in a total of 1,180 paraphrased property descriptions. We then evaluate whether \tool can still generate executable properties from these paraphrases.

\begin{figure}[t]
    \centering
    \begin{minipage}[t]{0.4\textwidth} 
        \centering
        \includegraphics[width=\textwidth]{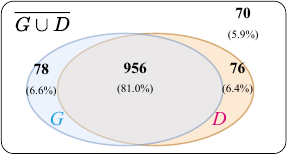}
        \caption{Venn diagrams of executable property generation. \textit{G}: GPT-4o; \textit{D}: DeepSeek-V3.}
        \label{fig:cp5_rq2_compare}
    \end{minipage}
    \hfill
    \begin{minipage}[t]{0.58\textwidth} 
        \centering
        \includegraphics[width=\textwidth]{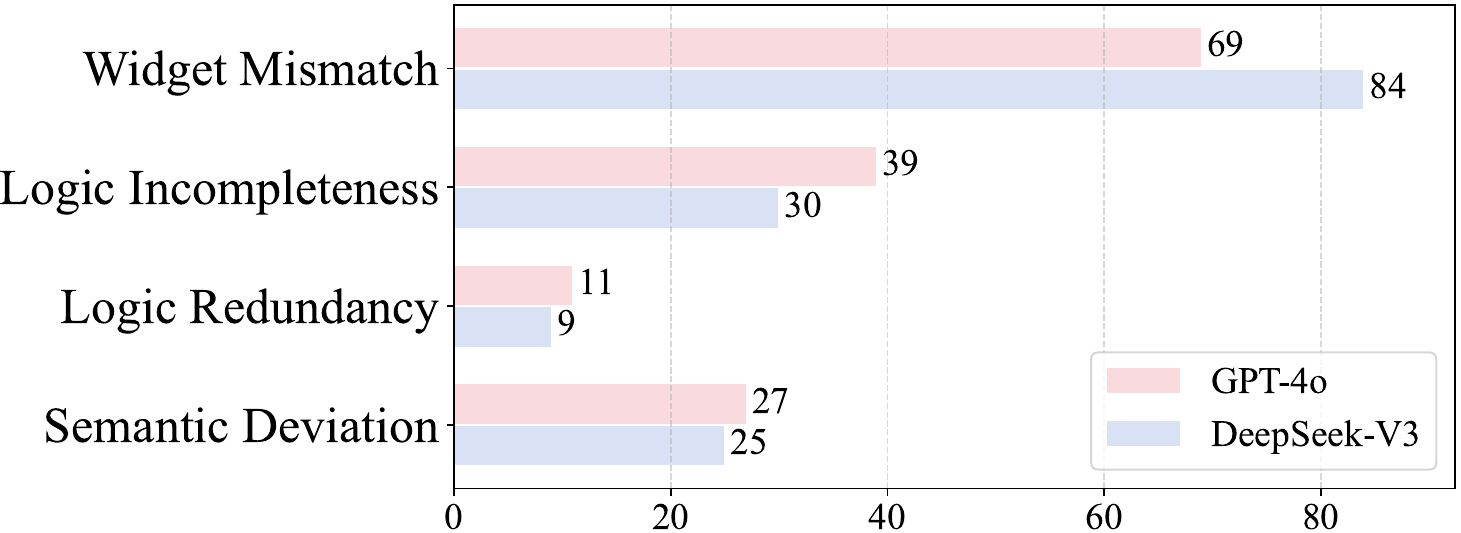}
        
        \caption{Failure symptom distribution.}
        \label{fig:cp5_rq2_sympton}
    \end{minipage}
\end{figure}



\noindent{\textbf{Evaluation results.}}
As shown in Fig.~\ref{fig:cp5_rq2_compare}, GPT-4o and DeepSeek-V3 generated 1,034 and 1,032 correct executable properties, respectively, achieving accuracy rates of 87.6\% and 87.5\%. Specifically, both models successfully generated correct executable properties for 81.0\% (956/1,180) of the descriptions, demonstrating that \tool is robust to variations in natural language expressions. 
Among the remaining cases, 6.6\% (78/1,180) of the descriptions were handled correctly by GPT-4o but not by DeepSeek-V3, while 6.4\% (76/1,180) were correctly processed by DeepSeek-V3 but not by GPT-4o. In 5.9\% (70/1,180) of the cases, both models failed to generate correct executable properties. 

To further understand the effectiveness of \tool in executable property generation, we analyzed the symptoms of the failure cases. Finally, we identified four main categories of failure symptoms.

Fig.~\ref{fig:cp5_rq2_sympton} presents the distribution of failure symptoms in the incorrect executable properties generated by \tool with two LLMs. The results reveal that UI widget mismatch is the most frequent error type, accounting for 47.3\%(69/146) and 56.8\%(84/148) of the total failures in executable properties generated by GPT-4o and DeepSeek-V3, respectively. This indicates that while LLMs demonstrate a general understanding of UI semantics, they still struggle to precisely align UI descriptions with their corresponding identifiers in some cases. Incomplete code is the second most common symptom, responsible for 26.7\%(39/146) and 20.3\%(30/148) of the errors. This is followed by other logic errors (18.5\%(27/146) and 16.9\%(25/148)), which typically involve incorrect API usage or faulty postcondition assertions. Redundant code ranks fourth, contributing 7.6\%(11/146) and 6.1\%(9/148) of the failures.
The similar distribution of errors across GPT-4o and DeepSeek-V3 demonstrates that \tool performs robustly across different LLMs, while also revealing typical challenges in executable property generation.

\begin{figure*}[t]
    \centering
    \includegraphics[width=\textwidth]{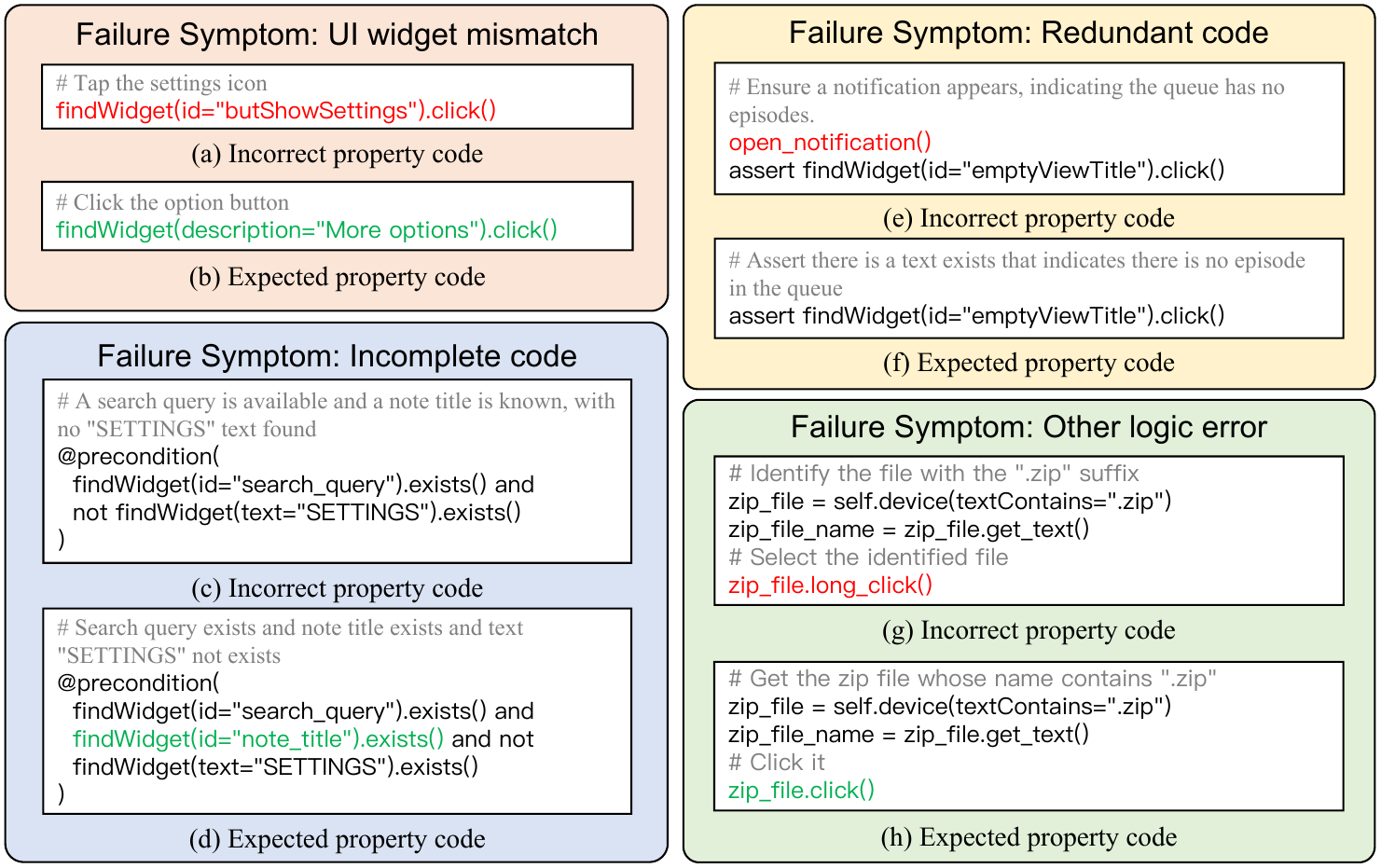} 
    \caption{Illustrative examples of failure symptoms. Incorrect and correct implementations are marked in red and green, respectively, with property descriptions in gray (code snippets are simplified for clarity).} 
    \label{fig:symptom_example} 
\end{figure*}

\begin{itemize}[leftmargin=*]
    \item\textbf{Widget Mismatch}. Widget mismatch occurs when the generated executable properties fail to match the intended UI widget. As shown in Fig.~\ref{fig:symptom_example}(a),  an example of incorrect executable properties generated by GPT-4o for the app AntennaPod~\cite{AntennaPod}. The expected UI event is to click the option button, but the UI widget identifier highlighted in red fails to match the target widget. In contrast, Fig.~\ref{fig:symptom_example}(b) shows the expected correct executable properties, where the UI widget identifier accurately matches the target UI widget.

    \item \textbf{Logic Incompleteness}. Logic incompleteness refers to generated executable properties that omit statements needed. This manifests primarily as (1) incomplete preconditions/postconditions, (2) missing conditional branches in interaction logic. Such incompleteness may lead to two consequences during execution: (1) execution failures during the property checking, (2) false positives in test results. Fig.~\ref{fig:symptom_example}(c) shows an incomplete code example, which was generated by DeepSeek-V3 for the app OmniNotes~\cite{omninotes}. The precondition fails to verify the existence of the "note title" UI widget, causing the testing framework to raise an exception while executing the interaction scenario when the widget is absent. Fig.~\ref{fig:symptom_example}(d) shows correct implementation, which includes the necessary existence check for the UI widget ( highlighted in green).

    \item \textbf{Logic Redundancy}. Logic Redundancy refers to the generated code containing unnecessary statements, primarily manifested as redundant UI events (\eg, \code{click}) in interaction logic. Such errors typically lead to execution failures during property checking. Fig.~\ref{fig:symptom_example}(e) shows an example, which was generated by GPT-4o for the app AntennaPod. The last line (highlighted in red) introduces an unnecessary event to open the notification page. Then, since the GUI state changes to the notification page, the assertion statement will fail during the property checking.
    Fig.~\ref{fig:symptom_example}(f) shows the expected executable properties.

    \item \textbf{Semantic Deviation}. This refers to other semantic errors, such as incorrect API usage or assertions in postconditions. These types of errors often result in property execution failures or inaccurate test results, including false positives and false negatives. Fig.~\ref{fig:symptom_example}(g) shows an example, which was generated by DeepSeek-V3 for the app Amaze~\cite{AmazeFileManager}. The implementation (highlighted in red) tries to long-click a file. However, the property description indicates that it should click the file. Fig.~\ref{fig:symptom_example}(h) shows the expected executable properties. 
\end{itemize}

\textit{Ablation study.}
We remove the widget functionality annotations from the prompt. The results show that  \tool generated 833 and 822 correct executable properties on GPT-4o and DeekSeek-V3, corresponding to accuracy rates of 70.6\% and 69.7\%, respectively. In comparison, the full setting achieved 87.6\% and 87.5\% accuracy. This represents a performance drop of 17.0 and 17.8 percentage points, respectively, highlighting the critical role of functionality annotations in guiding the LLM to generate correct executable properties.

\section{Discussion and Lessons}
\paragraph{Generality of our work}
First, the evaluation results demonstrate that our approach is effective in translating natural language property descriptions into executable properties for 124 properties. These 124 properties are collected from the existing dataset~\cite{kea} that contains 124 real functional bugs across 8 popular apps. It is interesting to further understand the generality of the approach on a larger set of apps. 
Second, our work's core methodology is translating informal specifications into formal properties in mobile apps, and it can be extended naturally to other similar applications, \eg, web applications (which also have UI widgets and user interactions) and formal specification generation in program verification. For example, many verification tools (\eg, Dafny~\cite{Dafny}) require formal pre/post-conditions or invariants, which share structural similarities with PBT properties.

\paragraph{Applying LLMs to PBT of mobile apps.}
Although existing PBT frameworks are effective, specifying properties requires significant manual effort and deep familiarity with UI structures and framework-specific APIs.
Allowing testers to write properties in structured natural language shifts this burden. In practice, this change enables testers to focus on what the app should do, rather than how to encode it, substantially lowering the barrier to using PBT.

\paragraph{Importance of UI semantic grounding.}
A central lesson is that UI semantic grounding is essential. Without widget functionality annotations, LLMs frequently select incorrect UI widgets, even when the generated code logic is otherwise correct. Our ablation results confirm that raw widget identifiers alone are insufficient in real-world apps, where identifiers are often ambiguous or poorly named.

\paragraph{Designing natural language as an effective specification interface} 
We found that natural language property descriptions should not be treated as completely free-form input. In practice, adopting a lightweight structure (precondition–interaction–postcondition) significantly reduces ambiguity and improves generation quality. This indicates that natural language specifications function as an interface between humans and LLMs, and even minimal structural constraints can greatly enhance reliability without harming usability.

\paragraph{What the failures reveal.}
Most failures arise from UI widget mismatch, not from incorrect control flow or API usage. This indicates that current LLMs can generally synthesize reasonable test logic, but still struggle with fine-grained UI disambiguation when multiple widgets have similar semantics.
In addition, paraphrased descriptions sometimes omit implicit constraints, leading to incomplete preconditions or redundant actions. This suggests that LLM-based property synthesis is robust to linguistic variation, but sensitive to semantic underspecification.

\paragraph{Complementing rather than replacing existing PBT frameworks}
Rather than replacing existing property-based testing frameworks, \tool complements them by addressing one of their most labor-intensive stages: executable property authoring. Since iPBT does not modify the execution or input generation mechanisms of PBT, it can be integrated into existing workflows with minimal disruption. This design choice proved important for maintaining practicality.


\paragraph{Threats to Validity.}
Our work may suffer from some threats to validity.
\textbf{\textit{First}}, the properties in our experiment may not fully represent those in real-world apps. To mitigate this threat,  we selected properties from the Kea dataset~\cite{kea}, where each property is derived from a real historical bug and covers important app functionalities. In the future, we will include a broader range of properties from industrial apps. Also, as we utilize dynamic exploration to collect the UI widget identifier list, it may not capture all the UI widgets in the app. This insufficient exploration is also identified as a common challenge of input generation in testing apps~\cite{behrang2020seven,su_stoat_2017}.
\textbf{\textit{Second}}, the natural language property descriptions may differ from how practitioners describe properties in practice.
To mitigate this, the initial descriptions were authored by an experienced property-based testing expert and carefully reviewed by all co-authors.
In addition, RQ2 evaluates descriptions written by 10 real users, and RQ3 further assesses robustness under diverse paraphrased descriptions.
\textbf{\textit{Third}}, our evaluation considers only two LLMs (GPT-4o and DeepSeek-V3), which may limit generalizability. We selected them as representative closed-source and open-source state-of-the-art models, and future advances in LLMs are expected to further improve performance.
\textbf{\textit{Finally}}, there is a potential concern that the models may have seen the evaluated properties during training.
This threat is unlikely, as both models were trained before the release of the Kea dataset, and all property descriptions were manually created and have not been publicly available.
To further avoid bias, we used a different LLM (Llama-3.1) for paraphrasing in the robustness evaluation.


\section{Related Work}

\noindent\textbf{LLM for SE tasks.}
The rapid advancement of LLMs has spurred significant research into their application across various software engineering domains. In code completion~\cite{nijkamp2022conversational,ding2023crosscodeeval,zhang2023repocoder}, LLMs have demonstrated strong capabilities in providing context-aware suggestions and recommendations. The domain of program repair~\cite{zhang2024survey,xia2023keep} has also benefited from LLMs' ability to understand and fix bugs in code.  Moreover, comprehensive studies~\cite{zhang2023survey,hou2024large} have systematically evaluated LLMs' capabilities across multiple software engineering tasks, understanding the applications, effects, and limitations. Our work specifically addresses the area of executable property generation for PBT.


\noindent\textbf{Automated test generation.}
The research on automated test generation can be categorized into two types: traditional and deep learning-based approaches. Traditional approaches include techniques such as fuzzing~\cite{afl}, symbolic execution~\cite{godefroid2005dart,sen2005cute,tillmann2014transferring}, search-based~\cite{fraser2011evosuite,pacheco2007randoop}. These approaches primarily aim to achieve high test coverage. 
However, these approaches often struggle to generate assertions~\cite{panichella2020revisiting,shamshiri2015automated}.

Deep learning-based approaches leverage the capabilities of pre-trained language models to generate tests from code snippets~\cite{tufano2020unit,lahiri2022interactive,yuan2024evaluating,yang2024evaluation}. For example, \textsc{AthenaTest}~\cite{tufano2020unit} leverages a transformer model, \textsc{BART}~\cite{lewis2019bart}, to generate unit test cases based on the given method input. In recent years, LLM-based approaches have shown promising results in test generation~\cite{yuan2024evaluating,jiang2024towards,deng2023large,lahiri2022interactive}. \textsc{TiCoder}~\cite{lahiri2022interactive} leverages LLMs to formalize user intent into tests, and 
\textsc{ChatTester}~\cite{yuan2024evaluating} can generate unit tests through interactive conversations with LLMs.
While these approaches focus on generating unit test cases, some recent work has moved closer to property-based testing. Endres \etal \cite{endres2024can} conduct a study to evaluate the LLM's capability of generating postconditions for individual functions based on the function comments. Vikram \etal ~\cite{vikram2023can} propose an approach to leverage LLM for generating property-based tests from specifications for Python libraries. Liu \etal ~\cite{liu2024propertygpt} target properties for smart contracts. In contrast, our approach focuses on mobile apps, generating executable properties (including preconditions, interaction scenarios, and postconditions) from natural language descriptions to capture real app behavior while reducing manual effort.

In mobile app testing, recent work like Kea~\cite{kea}, PBFDroid~\cite{pbfdroid}, and PDTDroid~\cite{sun2024property} has demonstrated the effectiveness of PBT in detecting functional and privacy bugs. However, these frameworks primarily focus on the execution and input generation phases, assuming the existence of high-quality executable properties. Consequently, they still require significant manual effort and domain expertise to write executable properties. Our work complements these approaches by automating the executable property generation.

Some work leverages LLM to analyze GUI pages during the dynamic exploration to find data inconsistency bugs~\cite{hu2024autoconsis}, functional bugs~\cite{liu2024seeingbelievingvisiondrivennoncrash}, or inconsistencies between app design and implementation~\cite{liu2025guipilot}. Our work differs from these approaches in its focus: rather than detecting bugs directly, we generate executable properties that can be used by existing PBT frameworks. Recently, different agents have been proposed to automatically perform tasks on mobile apps~\cite{wen2024autodroid,wen2023droidbot,zhang2025appagent,wang2024mobile,qin2025ui}. These works focus on executing user-specified tasks, whereas our work centers on generating executable properties to guide property-based testing.

\noindent\textbf{UI widget understanding.}
Various approaches try to understand the widget from different perspective~\cite{xiao2019iconintent,li2023you,malviya2023fine,liu2024unblind,xi2019deepintent,mahmud2025combining}. For example, \textsc{IconIntent}~\cite{xiao2019iconintent} leverages program analysis and computer vision techniques to identify sensitive UI widgets in Android apps. \textsc{DroidGem}~\cite{malviya2023fine} leverages deep neural networks to predict the permissions behind the UI widgets. \textsc{HintDroid}~\cite{liu2024unblind} aims to generate hint-text of the UI widget to improve the accessibilty for low-vision users. In contrast, our work focuses on constructing UI widget context for matching the property description with widget identifiers in PBT.

\section{Conclusion}
In this work, we presented a novel approach to automatically generate executable properties from natural language property descriptions. 
Our approach lowers the manual effort and technical expertise required for property-based testing of mobile apps.
In detail, we first construct the enriched widget context by extracting GUI information and leveraging MLLMs to generate functionality annotations of widgets. Then, we employ in-context learning to guide LLMs in generating executable properties with the carefully designed prompt.
Evaluation results show that our approach achieves 95.2\% accuracy on original property properties, maintains over 87\% accuracy on paraphrased variations, and substantial reductions in manual effort according to our user study. These results demonstrate that \tool makes PBT more practical and accessible for mobile app testing.



\bibliographystyle{ACM-Reference-Format}
\bibliography{ref}

\end{document}